%Paper: hep-th/9408105
%From: Alexander Laufer <alex@spock.physik.uni-konstanz.de>
%Date: Thu, 18 Aug 94 17:24:02 +0100
%Date (revised): Fri, 19 Aug 94 18:32:49 +0100
%Date (revised): Tue, 30 Aug 94 14:37:04 +0100

\magnification=1200
\abovedisplayskip=12pt plus 1pt minus 2pt
\overfullrule=0pt
\parindent=0pt
\advance\baselineskip by 3pt\
\font\titelf=cmbx10 scaled \magstep2
\font\gross=cmbx10 scaled \magstep1

\global\newcount\parnum
\global\newcount\glnum
\global\newcount\leerraum \leerraum=30
\def\newpar#1{\global\advance\parnum by 1
              \global\glnum = 0
              \bigbreak\bigbreak
               {\gross \the\parnum . \quad #1 }
              \bigskip}
%%
%  numbers of equations
%%
%
\def\gln{\global\advance\glnum by 1
           (\the\parnum .\the\glnum )}
\def\glnm{\global\advance\glnum by 1
          \eqno{(\the\parnum .\the\glnum )} }
\def\foono{\baselineskip=9pt  \sevenrm}
\def\abgl{\quad = \quad }
\def\tabgl{\quad &=\vphantom{\hbox to \leerraum pt {\hfill}} \quad }

\def\p{\; +\; }
\def\m{\; -\; }

\def\spur#1{\, \hbox{Tr} \, #1 \, }
\def\bruch#1#2{ \raise3pt \hbox {$ #1 $} / \lower3pt\hbox{$ #2 $}}
\def\quot#1#2{ \textstyle \, {{ #1}\over {#2}} \, }
\def\cent{\centerline}
\def\ssk{\smallskip}

\def\msk{\medskip}
\def\vsk{\vskip}

\topinsert \vskip .5in \endinsert
\centerline{\titelf THE EXPONENTIAL MAP FOR THE }
\medskip
\centerline{\titelf CONFORMAL GROUP 0(2,4) }
\vskip .5in
\cent{\bf A. O. Barut, $\;$ J. R. Zeni$\,$ {\dag }\footnote{*}{\foono
permanent address, Depto Ciencias Naturais, FUNREI, S\~ao Jo\~ao del
Rei, M.G., BRAZIL, 36.300,\hfil\break} $\;$ and $\;$ A.
 Laufer$\,$\footnote{**}{\foono permanent address, Physics
Department,
 University of Konstanz, mailbox 5560 M678, 78434 KONSTANZ,
 \hfill\break GERMANY, e-mail alex@spock.physik.uni-konstanz.de
 \hfill\break }\footnote{\dag}{\foono work partially supported by
 CAPES/BRAZIL and DAAD/GERMANY}}

\cent{Physics Department, University  of Colorado, Boulder, CO,
U.S.A.,
80.309-390}

\vsk 1in

{\bf Abstract} : We present a general method to obtain a  closed,
finite formula
for the exponential map from the Lie algebra to the Lie group, for
the
defining representation of the orthogonal groups.  Our method is
based
on the Hamilton-Cayley theorem and some special properties of the
generators of the orthogonal group, and is also independent of the
metric. We present an explicit formula for the exponential of
generators of the $SO_+(p,q)$ groups, with $p+q = 6$, in particular
we
are dealing with the conformal group $SO_+(2,4)$, which is
homomorphic
to the $SU(2,2)$ group. This result is needed in the generalization
of
U(1) gauge transformations to spin gauge transformations, where the
exponential plays an essential role. We also present some new
expressions for the coefficients of the secular equation of a matrix.

\vfil\eject

\newpar{INTRODUCTION}

The well known important formulae for the groups SU(2) and SO(3),
$$
 \eqalign{ e^{i \,{\theta \over 2}\; \sigma . \vec n} & \abgl \cos
 \theta /2 \; I_2 \p i \, \sigma . \vec n \;\sin \theta /2 \cr
 e^{\theta \,{\cal L}_j \;  n_j }& \abgl I_3 \p {\cal L}_j  \, n_j \;
 \sin \theta \p ({\cal L}_j \,n_j)^2 \;(1 \m \cos \theta) \cr } \glnm
$$
has been recently generalized to the group SL(2,C) and its
homomorphic
group $SO_+(1,3)$ [Zeni and Rodrigues, 90 and 92], but no such
formulae
seems to exist
 in the literature for the group $SO_+(2,4)$. The purpose of the
present work is to fill this gap,  presenting an explicit finite
formula for the series of the exponential of a g-skew-symmetric
matrix,
eq.(4.8), which represents the infinitesimal generators of the
orthogonal group $SO_+(2,4)$.
\par
The group $SO_+(2,4)$, or its covering $SU(2,2)$, appears in several
different contexts in theoretical physics. For instance, it is the
invariance group of the bilinear invariants in the Dirac theory of
electron. It is also homomorphic to the relativistic conformal group,
the largest group that leaves the Maxwell equations invariants
[Bateman],[Cunningham], or in other words, it is the largest group
which preserve the light cone of the Minkowski space-time [Gursey],
[Barut, 71], for a good review see [Fulton et al].  Other
applications
are found in the study of dynamical groups [Barut, 72].
\par
More recently, the group $SU(2,2)$ or its subgroups has been used in
spin gauge theories, in an attempt to generalize the minimal coupling
and to unify electrodynamics and gravitation [Dehnen et al.],
[Chisholm
and Farewell], [Liu], [Barut and McEwan], and in conformally
compactified space-times [Barut et al.].
\par
The mathematical problem can be stated basically as the summing up of
the exponential series for the matrix representing the generators of
orthogonal groups. The Hamilton-Cayley theorem plays a key role to
solve this problem, since it gives a recurrence relation between the
powers of the matrix, and so we can transform the matrix series into
real numbers series.
\par
The exponential map, as well other kinds of parametrizations for the
group elements of unitary groups, in particular $SU(3)$ and $SU(4)$,
deserves more attention in the literature, since the unitary groups
play an important role in Quantum Mechanics and particle physics
(see
[Barnes and Delbourgo] and references therein). In particular,
the work of [Bincer] on unitary groups, which  presents a
parametrization of the exponential map through a set of orthonormal
vectors obtained by considering the diagonal form of
the generators. The work of [Bincer] is related to the Jordan form
method of constructing
the exponential, since the latter provides a parametrization of
the exponential by the eigenvectors of the matrix [Faria-Rosa and
Shimabukubo].
\par
There are several articles in literature concerned   with the
exponential of an arbitrary matrix, for instance the work
of [Moler and van Loan] has a comprehensive review of methods,
analytical and numerical, for dealing with the exponential of an
arbitrary matrix as well as an extensive list of references.
\par
We remark that besides the exponential map there are other possible
parametrizations of group elements using the Lie algebra, see for
example [Lounesto] for the Cayley map.  However, the exponential map
deserves special attention due to its relationship with systems of
differential equations as discussed below.
\par
The symmetry properties of the matrices representing the generators
of
orthogonal groups are also helpful as they allow us to separate the
series into even and odd powers. It is a remarkable characteristic of
orthogonal groups that either even or odd powers occur in the
Hamilton-Cayley theorem, which amounts to a great simplification for
the sum of the exponential series.  These facts are discussed in the
Section 2 and 3, where general recurrence relations are obtained for
the powers of the generators of orthogonal groups.
\par
An important step in summing  up the series of a matrix is to
consider
the eigenvalues of the matrix instead of the coefficients which
appear
in the Hamilton-Cayley theorem.  We expect the series for the
exponential of a matrix to be expressed by means of elementary
functions of eigenvalues of the matrix, because if we consider the
solutions of a system of first order differential equations
$$
  {d \, X \over d\, t} \abgl H \, X \glnm
$$
The solutions are given by the exponential of $H$ parametrized by t,
i.e. $ X(t) \abgl e^{H t} \, X_0 $ (see [Magnus] and [Fer] for the
cases where $H$ is a function of time). On the other hand, we can
express the components of the vector $X(t)$ as exponential (scalar)
functions of the eigenvalues, as given below
$$
 X_j(t) \abgl  e^{\lambda _k \, t} \, C_{jk} \glnm
$$
where the $C_{jk}$ are chosen to fit the initial value $X_0$.
\par
Therefore, if we compare both solutions, it is obvious that the
matrix
elements of the exponential of a matrix must be elementary functions
of
the eigenvalues.  Another way to see that the above assertion holds
is
to look  at the Jordan form of the matrix $H$ [Faria-Rosa and
Shimabukubo].
\par
In Section 4 we derive an explicit finite formula for the exponential
of a matrix representing the generator of $SO_+(p,q)$ group, with
$p+q=6$. In order to close the series easily we have used
the discriminant related to the secular equation. The series for the
exponential is written as a product of elementary functions of the
eigenvalues by the first few powers of $H$. We remark that the series
is obtained in an intrinsic way, without explicit reference to the
matrix elements.
\par
In Section 5 we introduce a matrix representation for the generators
of
the $SO(2,4)$ and some formulae which simplify the previous results,
especially a new expression for the odd powers. We specialize the
previous result for the Lorentz group recovering the result of [Zeni
and Rodrigues, 90] and also eq.(1.1).
\par
In the Appendix we are going to discuss a new method to obtain the
coefficients of the secular equation from the trace of the powers of
the matrix.
\newpar{THE GENERATORS OF ORTHOGONAL GROUPS}
The matrices related to the defining representation of the orthogonal
groups, $O(p,q)$,   are defined by the following condition:
$$ A^t \, g \, A \abgl g, \qquad \hbox{or }, \qquad g \, A^t \, g
\abgl
A^{-1} \glnm $$
where g is a diagonal matrix with $p$ entries equal to $+1$ and $q$
entries equal to $-1$,  and the superscript t indicates the
transposed
matrix.
\par
The connected component of the identity of
the orthogonal groups will be hereafter indicated by $SO_+(p,q)$. In
what follows we are concerned with those transformations $A \in
SO_+(p,q)$ that can be written as $e^H$ where $H$ is called an
infinitesimal generator of the group, i.e. an element of the Lie
algebra [Miller], [Barut and Raczka].
Eq.(2.1) show that the generators of the orthogonal group are
given by,
$$ g \, H^t \, g \abgl - \, H \glnm $$
since we have that $g e^{H^t} g = e ^{g H^t g}$.
\par
The generators of the orthogonal group will be called here
g-skew-symmetric null diagonal matrices. The number of independent
real
parameters is $(n^2 - n)/2$ , which corresponds to the number of
elements in the upper (or lower) triangular matrix.
\bigbreak
{\bf Matrix Symmetry of the Powers of the Generators }
\msk
{\bf Lemma}: \ The odd powers of H are again g-skew-symmetric null
diagonal matrices. On the other hand, the even powers of H are
g-symmetric matrices, i.e.,
$$
H^{2n} \abgl g \, (H^t)^{2n} \, g, \qquad \hbox{while} \qquad H^{2n +
1} \abgl \, - \, g \, (H^t)^{2n + 1} \, g \glnm
$$
This result simplifies  the task of finding a finite, closed form for
the exponential, since it shows that we can work separately with the
series of even and odd powers,
$$ \eqalign{ A \abgl e^H  \tabgl \bruch{1}{2}\, (A \p g \, A^t \, g)
\p
   \bruch{1}{2}\, (A \m g \, A^t \, g)  \cr  \noalign{\vskip 10pt}
   \tabgl \; \sum^{\infty}_{n=0} \; {H^{2n} \over {2n!}} \quad +
\quad
\sum^{\infty}_{n=0}
 \; {H^{2n +1}\over {(2n +1)!}} \cr } $$
\msk
\par
We will show that the recurrence relations for both series of even
and
odd powers are similar, i.e., we can deduce one from another.
\bigbreak
{\bf Parity of Secular Equation}
\msk
We consider now the secular equation for g-skew-symmetric matrices
$$ \det(H \m \lambda \, I_n) \abgl 0 \glnm $$
{\bf Lemma}: Let H be a n$\times$n g-skew-symmetric matrix. If n is
odd
(even) only odd (even) powers of the eigenvalues are present in the
secular equation, i.e., ([Turnbull] presents a proof for the
euclidean
case, but the result holds for every metric),
$$ \det(H \m \lambda) \abgl (-1)^n \, \biggl( \lambda^{n} \p C_2 \,
    \lambda ^{n - 2} \p C_4 \, \lambda ^{n -4} + \ldots + C_{n - x}
\,
    \lambda ^x \biggr)\glnm
$$
where $x = 0$ if n is even and $x = 1 $ if n is odd.
\ssk
The lemma follows from the identity below which can be proved by
using
the g-skew-symmetry of $H$, eq.(2.2),
$$ \det ( H \m \lambda) \abgl (-1)^n \det ( H \p \lambda) \glnm $$
Thus it is clear that the determinant has a defined parity under the
change of $\lambda $ to -$\lambda $.
\newpar{THE HAMILTON-CAYLEY THEOREM}
The Hamilton-Cayley theorem guarantees that a matrix satisfies  an
matrix equation as its secular equation, i.e., for the generators of
the orthogonal groups we have
$$  H^n \; + \; C_2 \, H^{n-2} \; + \; C_4 \, H^{n-4} \; + \; \ldots
\;
+ \; C_{n -x} \, H ^x \; \abgl \; 0 \glnm $$
where the $C's$ are the coefficients of $\lambda$ in the secular
equation, eq.(2.5).
\ssk
According to equation (3.1) we can express $H^n$, in the case of even
n, in terms of the following set of matrices $H^{n-2}, H^{n-4},
\ldots,
I_n$.  By iterating eq.(3.1) by $H^2$ we can express all even  powers
of H in terms of the same set of matrices as discussed below.
\par
Also, we can apply an analogous reasoning to express all the odd
powers
of $H_{n\times n}$, with $n$ even, by means of the set $H^{n-1},
H^{n-3}, ..., H$.
\bigbreak
{\bf Recurrence Relations for the Powers of the Generators }
\msk
In the following we consider $n\times n$ matrices with even n, i.e.,
generators of the groups $O(2m)$, $2m=n$, and consider the recurrence
relations for even powers. In the case of the groups $O(2n +1)$ the
recurrence relations shown below can be obtained in an analogous way.
\par
Let us change our previous notation and write the recurrence relation
resulting
 from the secular equation, eq.(3.1), as follows:
$$ H^n \abgl a_0 \, H^{n-2} \p b_0 \, H^{n-4} \p \ldots \p v_0 \, H^2
\p x_0 \, I_n \glnm $$
where we are considering that n is even and $I_n$ denotes the
$n\times n $ identity matrix.  In general we set ($k\ge 0$)
$$ H^{n+2k} \abgl a_k \, H^{n-2} \p b_k \, H^{n-4} \p \ldots + v_k \,
H^2 \p x_k \, I_n \glnm $$
We are going to determine a recurrence relation for the coefficients
$a, b, \ldots, x$ present in the previous equation. To do this we
iterate the equation for $H^{n+2k}$ multiplying it by $H^2$, and
after
substituting eq.(3.2) for $H^n$, we obtain,
$$ a_{k+1} \, = \, a_0 a_k + b_k, \quad b_{k+1} \, = \, a_k b_0 +
c_k,
\quad \ldots, \quad v_{k+1} \, = \, a_k v_0 + x_k, \quad x_{k+1} \, =
\,
   a_k x_0 \glnm $$
The recurrence relations are the same for the coefficients of even
and
odd powers, provided that we define for the odd powers the following
formula (n even)
$$ H^{n + 1 + 2k} \abgl a_k \, H^{n - 1} \p  b_k \, H^{n - 3} \p
\ldots
   \p v_k \, H^3 \p x_k \, H    \glnm
$$
A closer analysis of eq.(3.4) shows  to us that if we determine the
coefficient $a_k$ then all others coefficients will be determined.
The
key to this problem is that the last coefficient $x_k$ is expressed
by
means of $a_k$. Furthermore, we can get a recurrence relation for
$a_k$
only involving $a_j$, $j < k $, by substituting the recurrence
relations for $b_k$ in the expression for $a_k$.
\newpar{THE ORTHOGONAL GROUPS \quad $O(p,q)$, \quad $p+q = 6$.}
For the orthogonal groups $O(p,q)$,  where $p + q = 6$, the secular
equation is a third order polynomial in $H^2$
$$ H^6 \m a_0 \, H^4 \m b_0 \, H^2 \m c_0 \, I_6 \abgl 0  \glnm $$
The previous recurrence relations for the coefficients $a_k, b_k$,
and
$c_k$, eq.(3.4),  can now be put in a simpler form, which ensures
that
if we determine the coefficients $a_k$ the others are determined, as
mentioned before
$$ \eqalign{c_{k+1} &\abgl a_k \, c_0 \hskip 5.25 truecm k\geq 0 \cr
   \noalign{\vskip 3pt} b_{k+1} & \abgl a_k \, b_0 \p a_{k-1} \, c_0
 \hskip 3.48 truecm k\geq 1 \cr
   \noalign{\vskip 3pt} a_{k+1} & \abgl a_k \, a_0 \p a_{k-1} \, b_0
\p
 a_{k-2} \, c_0 \qquad \qquad
  k\geq 2 \cr } \glnm $$
\ssk
where the first values follow from eq.(3.4)
$$ a_1 \abgl a_0^2 \p b_0, \qquad b_1 \abgl a_0 \,b_0 \p c_0, \qquad
   a_2 \abgl a_1 \,a_0 \p a_0 \,b_0 \p c_0    \glnm
$$
Our goal is to get a general formula for the coefficients $a_k, b_k,
c_k $ from these recurrence relations in terms of the eigenvalues.
For
this purpose we express the coefficients  $a_0, b_0$, and $c_0$, that
appear in the secular equation, eq.(4.1), in terms of the
eigenvalues,
which are the roots of that equation.  We remark that the secular
equation is a third order equation in the square of the eigenvalues,
so
that the eigenvalues appear in pairs, that we will indicate by $\{\pm
x, \pm y, \pm z\}$. We are going to use only three eigenvalues, $\{
x,
y, z\}$.  Therefore we have
$$ \eqalign{ a_0 & \abgl x^2 \p y^2 \p z^2 \cr  \noalign{\vskip 3pt}
   b_0 & \abgl \m x^2 \,y^2 \m x^2 \,z^2 \m y^2 \,z^2 \cr
   \noalign{\vskip 3pt} c_0 & \abgl x^2 \,y^2 \,z^2 \cr } \glnm
$$
We shall also make use of the following multiplier, $w$, to simplify
the expression for the recurrence relations. $w$ is equal to  the
square root of the discriminant of the secular equation, and is given
by
$$ w \abgl (x^2 \m y^2)\, (z^2 \m x^2)\, (y^2 \m z^2) \glnm $$
Writing the above recurrence relation for $a_k$, eq.(4.2), by means
of
$x, y$ and $z$, and after multiplying by $w$, we find the following
final form of the recurrence relations
$$ w \,a_k \abgl  (x^2 - z^2)\; y^{2k +6} \p (z^2 - y^2) \; x^{2k +6}
\p (y^2 - x^2) \; z^{2k + 6} \glnm
$$
which can be proven by finite induction and holds for $k\geq 0$, in
spite of the fact that eq.(4.2) holds only for $k \geq 2$.
\par
Substituting the above expression for $w \,a_k$ into the recurrence
relation for $b_k$, eq.(4.2),  we get the following recurrence
relation
for the product $w\, b_k$
$$ w \,b_k \abgl (z^4 - x^4) \, y^{2k + 6} \p (y^4 - z^4) \, x^{2k
+6}
\p  \, (x^4 - y^4) \, z^{2k + 6} \glnm
$$
which holds again for $k\geq 0$.
\bigbreak
{\bf The Series for the Exponential \quad $O(p,q)$, \quad $p+q =6$}
\msk
We have seen that the series for the exponential of the generators of
orthogonal
groups can be conveniently divided into the series for even and odd
powers.  Moreover, each series can be expressed by means of only a
few
powers of the generator. Summing up our previous result we have
transformed the matrix series into  real numbers series for the
coefficients $a_k, b_k, c_k$, for which recurrence relations were
obtained in the last Section for the groups $O(p,q)$, $p+q = 6$.
\par
After substituting eq.(3.3) for $H^{6+2k}$  the series of even powers
of $H$,  multiplied by $w$, can be written as shown below
$$  I_6 \, \Biggr(w   +  \sum {w\,c_k \over (6 + 2k)!} \Biggr) \;
  + \; H^2 \, \Biggl( {w\over 2} + \sum {w\, b_k \over (6 +
  2k)!}\Biggr)  \; + \; H^4 \, \Biggl( {w \over 4!} + \sum {w \, a_k
  \over (6 + 2k)!}\Biggr)
$$
\ssk
where all the sums run from zero to infinity.

We can also write an analogous expression for the series of odd
powers.
\par
Considering the previous recurrence relations eq.(4.6), eq.(4.7) and
eq.(4.2), the series of even and odd powers can be summed easily and
the result for the exponential of $H$, multiplied by $w$, is given
by
$$ \eqalignno{ w \, & e^{H}  \abgl
%Series for the coefficient of m$\cdot$H$^2$
 \biggl[\left( y^4 \m z^4 \right) \, \cosh x \p \left( z^4 \m x^4
 \right) \, \cosh y \p \left( x^4 \m y^4 \right) \, \cosh z
\biggr]\;
H^2 & \gln \cr & &\cr
%Series for the coefficients of $H^4$
& \p\biggl[ \left( z^2 \m y^2 \right) \, \cosh x \p  \left( x^2 \m
z^2
\right)\, \cosh y \p  \left(y^2 \m x^2 \right) \, \cosh z \biggr]\;
H^4
& \cr & & \cr
%Series for the coefficient of  m$\cdot$identity
& \p \biggl[\left( x^2 \m z^2 \right) \, x^2 \, z^2 \, \cosh y \p
      \left( z^2 \m y^2 \right) \, y^2 \, z^2 \, \cosh x \p
      \left( y^2 \m x^2 \right) \, x^2 \, y^2 \, \cosh z  \biggr]\;
I_6
& \cr & & \cr
%Series for the odd powers of H
%Series for the coefficient of m$\cdot$H
&\p \biggl[\left( x^2 \m z^2 \right) \, x^2 \, z^2 \, { \sinh y \over
y}
\p \left( z^2 \m y^2 \right) \, y^2 \, z^2 \,  { \sinh x \over x} \p
\left( y^2 \m x^2 \right) \, x^2 \, y^2 \, { \sinh z \over z}
\biggr]\;
H & \cr & &  \cr
%Series for the coefficient of m$\cdot$H$^3$
&\p\biggl[\left( z^4 \m x^4 \right) \,  \, {\sinh y \over y}
\p \left( y^4 \m z^4 \right) \,  \, {\sinh x \over x}
\p \left( x^4 \m y^4 \right) \,  \, {\sinh z \over z}  \biggr]\; H^3
& \cr & & \cr
%Series for the coefficient of m$\cdot$H$^5$
& \p \biggl[\left(  x^2 \m z^2 \right) \, \, {\sinh y \over y}
\p \left( z^2 \m y^2 \right) \,  \, {\sinh x \over x}
\p \left( y^2 \m x^2 \right) \,  \, {\sinh z \over z} \biggr]\; H^5
& \cr  }  $$
\ssk
The above series are valid for all groups $SO_+(p,q)$, with $p+q=6$.
The number of possible imaginary eigenvalues distinguishes the
metrics
and so transforms some of the hyperbolic functions given above into
trigonometric functions. For example, all eigenvalues of the
generator
$H$ are imaginary for the group $O(6,0)$.
\par
In the next Section we are going to introduce some particular cases
of
the formula above.
\newpar{ THE $SO_+(2,4)$ GROUP }
At this point in the discussion we want to focus on the group
$SO_+(2,4)$.  A generic generator of the group $SO_+(2,4)$ can be
represented by the following matrix:
$$ H \abgl h_{ij} \, {\cal L}^{ij} \abgl \pmatrix{
                      0  &  e_1  &  e_2  &  e_3  & v_0 & a_0 \cr
                     e_1 &   0   &  b_3  & - b_2 & v_1 & a_1 \cr
                     e_2 & - b_3 &   0   &  b_1  & v_2 & a_2 \cr
                     e_3 &  b_2  & - b_1 &   0   & v_3 & a_3 \cr
                     v_0 & - v_1 & - v_2 & - v_3 &  0  &  c  \cr
                   - a_0 &  a_1  &  a_2  &  a_3  &  c  &  0  \cr}
\glnm
$$
\msk
where the metric is given by $g = \hbox{diag}(+,-,-,-,-,+)$.
$h_{ij}$,
$i\leq j$ , are the matrix elements of $H$, and ${\cal L}_{ij} $ is
the
standard basis for
the Lie algebra of $SO_+ (2,4)$ in the defining representation, i.e.,
they are matrices with only two non-zero g-skew-symmetric elements,
which are equal to $\pm1$, [Barut, 71].
\ssk
The form above for the matrix generator, and therefore also for the
metric, was chosen in such a way that makes it possible to establish
a
closer connection with the generators of $SU(2,2)$, which can be
represented by the Dirac matrices [Barut, 64], [Kihlberg and Muller].
For example, the components $v_{\mu}$ and $a_{\mu}$, $\mu \in [0,3]$,
are related to a vector and axial vector in the Dirac algebra. On the
other hand, we set the generator of $SO_+(2,4)$ in such a way that
one
of its subgroups, the Lorentz group $SO_+(1,3)$, has a priviledged
place; it corresponds to the $4\times 4$ block formed by the $e_j$
and
$b_j$, $j=1,2,3$. The last component of the generator above, $c$,
correspond to the generator of a chiral transformation in the Dirac
algebra, i.e.  a transformation generated by $\gamma_5$.
\par
The coefficients $C_k$ present in the secular equation can be written
as (see also  Appendix, eq.(A.4) and eq.(A.5)),
$$\eqalignno{ C_2 \tabgl    \; - \,\quot{1}{2} \;h_{ij} \, h_{ji}
\abgl
 - \quot{1}{2} \spur{H^2} &   \cr
\noalign{\vskip 5pt}
   C_4 \tabgl % - b_0  =  g_{ii} \; g_{jj} \; a_{ij}^2
-\quot{1}{2}\,  p_{ij}\, p_{ji} \abgl - \quot{1}{2}\spur{P^2} &
\gln\cr
\noalign{\vskip 5pt}
  C_6 \tabgl % - c_0  = \det H   =  \bigl[ h_{ij} \; a_{ij} \bigr]^2
 \det H  \abgl \bigl[ \quot{1}{6}\spur {H\cdot P} \bigr]^2 \abgl
\bigl[
    \quot{1}{6}\, h_{ij}\, p_{ji} \bigr]^2  & \cr  }
$$
where we sum over repeated indices. The elements $p_{ij}$ of the
matrix
$P$ are defined below.
\par
We remark  that the determinant of a skew-symmetric matrix is equal
to
the square of a polynomial which defines the pfaffian of the matrix.
\par
The elements $p_{ij}$ introduced in eq.(5.2) are, except for a sign,
the pfaffian of the matrices obtained from $H$ cutting the i-th and
j-th rows and also the i-th and j-th columns  [Turnbull]. A general
expression for the $p_{ji}$, in this order, with $i<j$ is
$$ p_{ji} \abgl h_{kl} \; h_{mn} \m h_{km} \; h_{ln} \p h_{kn} \;
h_{lm} \glnm
$$
where (ijklmn) is an even permutation of (123456). Otherwise we need
to
change the sign of the terms in the right hand. We are also
considering
in the above formula only $i<j$, for each $h_{ij}$.
\ssk
We can form a matrix with the pfaffians agregate $p_{ij}$, that will
be
called hereafter the {\it pfaffian} matrix associated with $H$, and
will be indicated by $P$.  We define the pfaffian matrix $P$ such
that
it has the same symmetry as $H$, i.e., $(g\,P)^t = - g \, P$.  The
name
pfaffian for the matrix $P$ is well justified, since we have
$$ P \, H \abgl H \, P \abgl \, \sqrt {\det H } \; I_6 \quad
  {\buildrel \rm def \over = }  \quad \hbox{pfaff H} \; I_6 \glnm
$$
The explicit expression for the pfaffian of the matrix $H$, eq.(5.1),
is given by
$$
C_6 \abgl \biggl(c \, \vec e \hbox{\bf .} \vec b \m v_0 \, \vec a
\hbox{\bf .}   \vec b \p a_0 \, \vec b \hbox{\bf .} \vec v \p \vec e
\hbox{\bf .} (\vec a \times \vec v) \biggr)^2
$$
We are using a short-cut notation, $\vec e = (e_1,e_2,e_3)$, and
$\vec v, \vec b$ and $\vec a$ are three dimensional vectors from
which
we can compute the dot {\bf .} and cross $\times$ product in the
usual
way.
\par
Moreover, it is remarkable that the series for the odd powers can be
rewritten by means of $H, H^3$ and $P$, avoiding the use of $H^5$.
The
advantage to work with the matrix $P$ instead of $H^5$ is clear,
since
$P$ is second order in $H$.
The above remark is based on the following equation that can be
verified directly
$$ H^5 \abgl a_0 \, H^3 \p  b_0 \, H \m i \, \sqrt {\, c_0} \; P
\glnm
$$
where $ i = \sqrt{-\,1}$ is the imaginary unit. We remark that
$c_0 = - \det H $, cf. eq.(4.1).
\par
The above formula, as well as also the formulae presented in the
Appendix, eq.(A.4) and eq.(A.5), were checked by a symbolic
mathematical program.
\par
The recurrence relation which comes from the Hamilton-Cayley theorem,
eq.(4.1), can be obtained from the above equation.  However the
converse is not true, i.e., we can not obtain the latter equation
from
the Hamilton-Cayley theorem if $\det H = 0$,  but the above equation
holds even when $\det H = 0$ .
\ssk
The series for the odd powers given in eq.(4.8) can be rewritten by
means of $H, H^3$ and $P$. After substituting the above expression
for
$H^5$ and using eq.(4.4) to replace $a_0, b_0$ and $c_o$ by x, y and
z,
we find
$$ \eqalignno{ w \, & \sum {H^{2i+1} \over (2i +1)!} \; = \;
%Alternative Series for the coefficient of $w H$
\biggl[\left( z^4 - x^4 \right) \,  y \, \sinh y
\p \left( y^4 - z^4 \right) \,  x \, \sinh x
\p \left( x^4 -  y^4 \right) \,  z \, \sinh z   \biggr]\; H  \cr
\noalign{\vskip 12pt}
%Alternative Series for the coefficient of $w P$
& \m i \biggl[\left( x^2 -  z^2 \right) \, x \, z \, \sinh y \p
  \left( z^2 -  y^2 \right) \, y \, z \,   \sinh x  \p \left( y^2 -
  x^2 \right)  \, x \, y \,  \sinh z  \biggr]\; P  \cr
\noalign{\vskip 12pt}
%Alternative Series for the coefficient of $w H^3$
& \p \biggl[\left(  x^2 - z^2 \right) \, y \,  \sinh y
\p \left( z^2 - y^2 \right) \,  x \, \sinh x
\p \left( y^2 - x^2 \right) \,  z \, \sinh z  \biggr]\; H^3
& \gln \cr }
$$
\msk
Now, we shall discuss as a special case how the series for the
Lorentz
group $SO_+(1,3)$ presented in [Zeni and Rodrigues, 90],  can be
obtained from the series for $SO_+(2,4)$.  Let us write $F$ for the
matrix $H$, eq.(5.1), when $a_{\mu}, v_{\mu}, c$ vanish. Therefore,
the
proper and ortochronous Lorentz transformations are generated by the
matrix $F$. In this case, two of the eigenvalues, $\{\pm z\}$, of $F$
vanish, since $\det F = 0$. Moreover, the product of the other two
eigenvalues, $x$ and $y$,  can be written as $\vec e \hbox{\bf .}
\vec
b = i \, x \, y $, since in this case $C_4 = (xy)^2 = - (\vec e {\bf
.}
\vec b)^2$.
\par
We also obtain a simpler recurrence relation for the powers of $F$,
instead of the Hamilton-Cayley, eq.(4.1), or eq.(5.5) [Zeni and
Rodrigues, 90], now we have
$$ F^3 \m (x^2 \p y^2) \, F \m i \, x \, y \; G \abgl 0 \glnm $$
where $G$ is the dual (Hodge) matrix obtained from $F$ by changing
$e_j \to b_j$ and $b_j \to - e_j$. The dual $G$ has the following
significant property
$$ F \, G \abgl G \, F \abgl i \, x \, y\; J_4 \abgl (\vec e {\bf . }
\vec b) \; J_4 \glnm
$$
where  $J_4 = diag(1,1,1,1,0,0)$.
\par
{}From the two equations above it is clear that we can express $F^4$
and
$F^3$ by means of $F^2$, $J_4$ and $F$, $G$, respectively.
\par
We remark that if we put $z=0$ in eq.(4.8), the coefficient of
identity, $I_6$, becomes equal to $w' = - x^2 y^2 (x^2 - y^2)$, the
value of $\omega$, eq.(4.5) when $z=0$. Also the coefficient of the
pfaffian matrix $P$, given by eq.(5.6), vanishes in this case.
\par
Substituting eq.(5.7) for  $F^3(=H^3)$ in eq.(5.6), and for $F^4$ in
eq.(4.8), we can write the series for the $6\times 6$ matrix $F$ by
means of $J_4, F, F^2, G$ as
$$ \eqalignno{ e^F \tabgl I_6 \m J_4 \p \Biggl({x \sinh x - y \sinh y
\over x^2 - y^2}\Biggr) F  \p \Biggl({y \sinh x - x \sinh y \over x^2
-
  y^2}\Biggr) G  \cr \noalign{\vskip 15pt} & \p \Biggl({\cosh x -
\cosh
y \over x^2 - y^2}\Biggr) F^2 \p \Biggl({x^2 \cosh y - y^2 \cosh x
\over x^2 - y^2}\Biggr) J_4 &
  \gln\cr }
$$
\msk
The series presented in [Zeni and Rodrigues, 90, eq.(6)] are obtained
from the equation above by changing $y \to - i y'$,  except for the
additive factor $I_6 - J_4$.
\par
As a further special case, the generators of SO(3) are given by the
matrix $F$ when $e_j = 0$, $j\in [1,3]$. The exponential of a
generator
of SO(3), eq.(1.1) , can be obtained from the formula above by
putting
$x = 0$, changing $y \to -i \, \theta $, $J_4 \to J_3 =
\hbox{diag}(0,1,1,1,0,0)$, and considering that $ F = \theta \, {\cal
L}_j \; n_j$.
\newpar{CONCLUSIONS}
In this article we presented a finite formula for the exponential of
the Lie algebra to the conformal group SO(2,4), which is homomorphic
to
the special unitary group SU(2,2). This latter group has been used in
spin gauge transformations, where the exponential can be used to
determine the explicit form of the transformation of the Dirac
matrices
[Barut and McEwan].

We plan to discuss the  related
result to the exponential map for the SU(2,2) group in a forthcoming
paper, and establish some connections with  the works already
existing
in the literature for the unitary groups $SU(n)$, see for example
[Barnes and Delbourgo] and [Bincer].

At this point, we recall that the use of other algebras, in
particular
Clifford algebras,  can simplify the discussion, as seen in
[Zeni and Rodrigues, 92], where the Clifford algebra of space-time
was
used to get the exponential map  to $Spin_+(1,3) \sim SL(2,C)$ in a
very simple way, establishing a straithforward generalization of the
first part of eq.(1.1),  this in turn is the exponential of a
pure quaternion ([Silva Leite] obtained the exponential of octonions
as
another possible generalization of a quaternion exponential).  We are
particularly concerned with the Clifford algebra generated by the
vector space $R^{2,4}$, since we have $Spin_+(2,4) \sim SU(2,2)$.
\par
Our approach to obtain the exponential can be used for every
orthogonal
group, since eq.(3.4) holds universally. From that equation it is
possible to obtain an explicit formula for the exponential map from
the
Lie algebra in the connected component to the identity of the group,
as
we have achieved in this article for the $SO_+(2,4)$ group. It is a
remarkable result, since the expoential map is usually presented only
in the
infinitesimal form and assumed to hold only in a neighborhood of the
identity [Barut and Razcka], [Miller].
\par
Also from the finite formula for the group element we can easily
discuss the group law and related subjects.  In particular, it makes
the Baker-Campbell-Hausdorff formula superfluous [Miller], since it
provides an exact solution for problems related to this series (see
[Zeni and Rodrigues, 92]
for a discussion related to the Lorentz group, SL(2,C)).
\bigbreak
\newpar{ACKNOWLEDGMENTS}
The authors are thankful to Prof. W.A. Rodrigues Jr., P. Lounesto,
and
H. Dehnen for %several discussions on the subject.  fruitful
discussions. J. R. Zeni and A. Laufer are thankful to Prof. A. O.
Barut
for his kind hospitality in Boulder. A. Laufer and J. R. Zeni are
also
grateful to DAAD/GERMANY and CAPES/BRAZIL, respectively, for the
fellowships that supported their stay in Boulder.
The authors are also thankful to the referees for calling their
attention to some references related to this paper..
\bigbreak
\bigbreak
{\bf APPENDIX: \quad Eigenvalues with power series expansion}
\msk
Here we present some expressions for the coefficients of the secular
equation for general matrices and then we specialize our formulae for
the generators of orthogonal groups. We remark that formulae for
these
coefficients are usually obtained by using the minors of the matrix
[Turnbull]. Our approach is different, and gives the coefficient by
means of the trace of powers of matrix.
\medskip
We use the following formula for calculating the determinant
[Miller]:
$$\eqalignno{%
\hbox{det}\, \left( B \m \lambda \, I_n \right) \tabgl
\left( -\lambda \right)^n \; \exp \left[ \spur{ \ln \left( I_n \m
  \quot{1}{\lambda} B \right)} \right]  & \cr \noalign{\vskip 15pt}
\tabgl \left( -\lambda \right)^n \; \sum_{m=0}^{\infty} \left( -1
\right)^m \, {{1}\over {m!}} \left[ \sum_{l=1}^{\infty} \, {{1}\over
{l
\lambda ^l}}
\spur{ B^l } \right]^m & (A.1) \cr}
$$
\ssk
where we make use of the series expansion for $\ln (1 + x)$ and the
exponential functions.
\par
{}From this formula we can see directly that if the odd powers of the
matrix
$B$ are traceless, e.g. the generators of orthogonal groups, we have
only odd or even powers of the eigenvalue in the secular equation,
according to $n$ is odd or even (cf. eq.(2.10)).
\par
Collecting terms of the same power in eigenvalue $\lambda $ we get:
$$
\det\, \left( B \m \lambda \, I_n \right) \abgl
 \left( -\lambda \right)^n \; \Bigg( 1\p \sum_{k=1}^{\infty}
 {{1}\over {\lambda^k }} \, C_k \Bigg) \eqno{(A.2)}
$$
Since the determinant of the secular equation has only positive
powers
in $\lambda$, the series in negative powers of $\lambda$ is actually
a
finite series, and we must have $C_k = 0$, for $k >n$. To gain an
understanding for this, we recall the Hamilton-Cayley theorem which
provides us with a relationship for the powers $k \geq n$ of a
matrix.
\par
The coefficients $ C_k $ can be written in the following way
$$\eqalignno{%
C_k \abgl {{(-1)^k } \over {k!}} & \bigl(\spur {B}\bigr)^k & (A.3)
\cr
\cr \p \sum_{m=1}^{k-1} \, & \sum_{p=1}^{m}  {{(-1)^m } \over {m!}}
\bigl(\spur {B} \bigr) ^{m-p} \sum_{l_1 =1}^{L_1} \ldots
\sum_{l_{p-1}=1}^{L_{p-1}} \, \left( \prod_{i=1}^{p-1} {{1}\over
{l_i}}
\spur{B^{l_i}} \right) \, {{1}\over { l_p }} \, \spur{B^{l_p}}  & \cr
\cr}
$$
\ssk
with $ L_j = k+j-m-1-{\displaystyle \sum_{i=1}^{j-1}}\, l_i $\ and \
$ l_p =  k+p-m -{\displaystyle \sum_{i=1}^{p-1}} \, l_i $
\bigskip
The first three coefficients for a generic
matrix $B_{n\times n}$, with $n\geq 3$ are
$$\eqalignno{%
C_{1} \tabgl \m \spur{B}   \cr \noalign{\vskip 8pt}
C_{2} \tabgl \quot{1}{2}  \left(\spur{B}\right) ^2 \m \quot{1}{2}
\spur{B^2}
 \cr  \noalign{\vskip 8pt} C_{3} \tabgl - \quot{1}{3} \spur {B^3} \p
\quot{1}{2}  \spur{B} \spur{B^2} \m
	\quot{1}{6} \left( \spur{B}\right)^3   & (A.4) \cr}
$$
\ssk
As mentioned before, if the odd powers of the matrix $B$ are
traceless,
the coefficients $C_k$, for $k$ odd,  vanish. In this case, the
formula
for the coefficients (eq.(A.3)) can be substantially simplified
because
the only non-vanishing terms are those with $m = p$ .
\par
We remark that the determinant of the matrix is given by the
coefficient $C_n$.  For example, for the generators of the $O(p,q)$
groups, with $p+q = 6$ we have that
$$ C_6 \abgl -\quot{1}{48} \left(8 \spur{B^6} \m 6
\spur{B^2}\spur{B^4}
       \p  \bigl(\spur{B^2}\bigr)^3  \right)  \eqno{(A.5)}
$$
Finally,we remark that if we represent a second rank tensor by a
matrix, the ``principal'' invariants of the tensor are just the
coefficients of the secular equation of the matrix [Landau and
Lifshitz].  The most common invariants related to a matrix are the
trace, the first coefficient of the secular equation, and the
determinant, i.e. the last coefficient in the secular equation.
However, all the coefficients present in the secular equation are
also
invariant. This becomes clear from the formula above, since these
coefficients are expressed by means of the trace.
\par
\newpar{REFERENCES}

K.J. BARNES and R. DELBOURGO (1972) J. Phys. {\bf A5}, 1043.

A.O. BARUT (1964) Phys. Rev., {\bf 135}, B839.

A.O. BARUT (1980) {\it Electrodynamics and Classical Theory of Fields
and Particles}, DOVER.

A.O. BARUT (1971) in {\it Lectures on Theoretical Physics}, vol.
XIII,
editors A.O. BARUT and W.E. BRITTIN, Colorado Univ. Press, Boulder,
pg.
3.

A.O. BARUT (1972) {\it Dynamical Groups and Generalized Symmetries in
Quantum Theory}, Univ. of Canterbury Press, Christchurch, New
Zealand.

A.O. BARUT, P. BUDINICH, J. NIEDERLE, R. RACZKA (1994) Found. of
Physics, at press.

A.O. BARUT and J. McEWAN (1984) Phys. Lett., {\bf 135}, 172.

A.O. BARUT and R. RACZKA (1986) {\it Theory of Group Representation
and
Applications}, second ed., World Scientific.

H. BATEMAN (1910) Proc. London Math. Soc., {\bf 8}, 288.

A. M. BINCER (1990) J. Math. Phys., {\bf 31}, 563.

J.S.R. CHISHOLM and R.S. FARWELL (1989) J. of Phys. A, {\bf 22},
1059.

E. CUNNINGHAM (1909) Proc. London Math. Soc., {\bf 8}, 77.

H. DEHNEN, F. GHABOUSSI (1986) Phys. Rev. D, {\bf 33}, 2205.

M.A. FARIA-ROSA and A. SHIMABUKUBO (1993) communication in {\it VI
Latin-American Symposium on Relativity and Gravitation}, ed. W.A.
Rodrigues and P.A. Letelier, World Scientific, in press.

F. FER (1958) Acad. Roy. Belg. Bull. Cl. Sci., {\bf 44}, 819.

T. FULTON, F. ROHRLICH and L. WITTEN (1962) Rev. Mod. Phys., {\bf
34},
462.

F. GHABOUSSI, H. DEHNEN and  M. ISRAELIT (1987) Phys. Rev. D, {\bf
35},
1189.

F. GURSEY (1956) Nuovo Cimento, {\bf 3}, 98.

A. KIHLBERG, V.F. MULLER and F. HALBWACHS (1966) Commun. Math Phys.,
{\bf 3}, 194.

L. LANDAU and E. LIFSHITZ (1951) {\it Classical Theory of Fields},
Addison-Wesley, reprinted from the Russian second edition. The
remarks
in the Section on field invariants that interest to us are missing in
latter editions.

Y. LIU (1992) Commun. Theor. Phys., {\bf 18}, 443.

P. LOUNESTO (1987) ``Cayley Transform, Outer Exponential and Spinor
Norm'', in {\it Supplemento ai Reendiconti del Circolo Matematico di
Palermo}, serie II, numero 16, Palermo (ITALY).

W. MAGNUS (1954) Comm. Pure and App. Math., {\bf 7}, 649.

W. MILLER Jr., (1972) {\it Symmetry Groups and Their Applications},
Academic Press.

C. MOLER and  C. VAN LOAN (1978) SIAM Review, {\bf 20}, 801.

F. SILVA LEITE (1993) Linear and Multilinear Algebra, {\bf 34}, 123.

H. W. TURNBULL (1960) {\it The Theory of Determinants, Matrices and
Invariants}, Dover.

J.R. ZENI and W.A. RODRIGUES (1990) Hadronic J., {\bf 13}, 317.
Eq.(6.d) is misprinted, it should read $g_2 (x,y) = (y \sinh x - x
\sinh y) /\, |z|^2$.

J.R. ZENI and W.A. RODRIGUES (1992) Int. J. Mod. Phys., {\bf A7},
1793.

\vfill\end